\documentclass[12pt]{article}

\usepackage{color}
\usepackage{amsmath}
\usepackage{amsfonts}
\usepackage{amssymb}
\usepackage{dsfont}
\usepackage{caption}
\usepackage{graphicx}
\usepackage{slashed}            
\usepackage{subcaption}
\usepackage{xspace}				
\usepackage{cite}
\usepackage{bm}
\let\originalleft\left
\let\originalright\right
\renewcommand{\left}{\mathopen{}\mathclose\bgroup\originalleft}
\renewcommand{\right}{\aftergroup\egroup\originalright}

\usepackage[english]{babel}
\usepackage{fancyhdr}
\usepackage{amsmath}
\usepackage{amssymb}
\usepackage{amsfonts}
\usepackage{psfrag}
\usepackage[applemac]{inputenc}
\usepackage{graphicx}

\renewcommand{\tilde}{\widetilde} 

\newcommand{\Tr}{\mathrm{Tr~}}

\newcommand{\beq}{\begin{eqnarray}}
\newcommand{\eeq}{\end{eqnarray}}

\newcommand{\bag}{\begin{align}}
\newcommand{\eag}{\end{align}}

\usepackage[hypertexnames=false]{hyperref}		

\begin{document}
\begin{titlepage}

\begin{center} 
{\huge \bf Anomalies at the \\ \vspace*{0.4cm} End of the Universe?} 
\end{center}
 
\begin{center} 
{\bf \  Jay Hubisz, Hanieh Moradipasha, and Prakriti Singh} 

 {\it Department of Physics, Syracuse University, Syracuse, NY  13244}

\vspace*{0.1cm}
{\tt  
 \href{mailto:jhubisz@syr.edu}{jhubisz@syr.edu},
 \href{mailto:hmoradip@syr.edu}{hmoradip@syr.edu},  
 \href{mailto:psingh39@syr.edu}{psingh39@syr.edu}}
\end{center}
\vglue 0.3truecm
\begin{abstract}
The spatial distribution of anomalies in extra dimensional theories on a slice of AdS leads to a puzzle where anomalies may not match across different geometries that are dual to various states of the same CFT.  We argue that consistency of the anomaly across such geometries necessitates the addition of topological terms or other bulk fermions to the theory that flow any IR localized consistent anomaly into the ``UV" region of these geometries if one wishes to have such a dual interpretation.
\end{abstract}

\end{titlepage}

\setcounter{equation}{0}
\setcounter{footnote}{0}
\setcounter{section}{0}

\section{Introduction}

Anomalies are a subtle consequence of quantum field theory with important theoretical and phenomenological implications.  They play a role in many aspects of the Standard Model (SM) and its extensions.  To name a few:  they are behind the resolution of the $\eta-\eta'$ mass puzzle~\cite{tHooft:1976rip,Witten:1979vv}, place rigid constraints on the matter content of the SM~\cite{Bouchiat:1972iq} and its extensions, play a critical role in the statement of the strong CP problem and its potential resolutions~\cite{Peccei:1977hh}, and tightly constrain the relationship between UV and IR effective degrees of freedom~\cite{tHooft:1979rat}.  See~\cite{Harvey:2005it, Reece:2023czb} for modern reviews.

Anomalies have an especially rich structure in extra-dimensional theories, which on their own serve as an arena for exploring physics beyond the SM (BSM).  Their potential relation to strongly coupled approximately conformal dynamics makes their study especially interesting, since they then offer a perturbative window into non-perturbative physics~\cite{Maldacena:1997re,Gubser:1998bc,Witten:1998qj}.  An especially popular class of constructions are the Randall-Sundrum I models, which are built on a slice of 5D AdS space that is truncated by both UV and IR branes~\cite{Randall:1999ee,Randall:1999vf}.   

In many such models, the SM fields are taken to propagate in the bulk, and the spectrum of modes we observe arises, in part, from the boundary conditions that are chosen for those bulk degrees of freedom. In the dual picture, the UV brane represents a high-energy cutoff of a CFT, with the cutoff rendered physical by the presence of fundamental fields coupled to the CFT (with such couplings typically explicitly violating the conformal symmetry).  A schematic Lagrangian representing this is:
\beq
{\mathcal L} = {\mathcal L}_0 + {\mathcal L}_\text{CFT} + \sum_{ij} g_{ij} O_0^i O_\text{CFT}^j,
\eeq
where $O^i_0$ and $O^j_\text{CFT}$ are operators of the fundamental and CFT sectors, respectively, and the $g_{ij}$ are coupling constants governing interactions that mix the two sectors~\cite{Rattazzi:2000hs,Arkani-Hamed:2000ijo,Perez-Victoria:2001lex}.

The UV brane position is associated with scales that characterize the fundamental sector, often taken to be the Planck scale associated with dynamical 4D gravity, though it could also represent a scale at which the theory exits a conformal window.  

The IR brane, in turn, represents the spontaneous breaking of the (approximate) conformal symmetry through CFT operators acquiring vacuum expectation values, generating a gap in the spectrum. The IR brane position is associated with the symmetry-breaking scale \(f\), which suppresses the interactions of the corresponding Goldstone boson (the radion/dilaton) that non-linearly realizes the spacetime/conformal symmetries  ~\cite{Goldberger:1999uk,Csaki:2000zn,Bellazzini:2012vz}.\footnote{An infinitesimally thin hard-wall IR brane is generally imagined to be only an effective description of the dynamics, with a more complete theory ``resolving" its microscopic description~\cite{Batell:2008zm}. A soft-wall construction constitutes a more natural mechanism for truncating the geometry~\cite{Karch:2006pv}.}

In light of these associations, the boundary conditions for various 5D degrees of freedom can then be interpreted similarly.  A 5D gauge field, for example, with Neumann (Dirichlet) boundary conditions on the UV brane, is dual to a global symmetry of the CFT that is gauged (not gauged).  If it is gauged, there is a gauge field in the fundamental sector coupled to a conserved current in the CFT.  On the IR brane, Neumann (Dirichlet) boundary conditions correspond to preservation (spontaneous breaking) of the global symmetry by the vacuum expectation values of operators that simultaneously break the CFT~\cite{Gherghetta:2000qt}.  In the case of Dirichlet conditions on both branes, the dual picture is of a spontaneously broken global symmetry, in which case the Goldstone theorem is satisfied by the presence of a massless fifth component of the gauge field~\cite{Contino:2003ve}.

For fermions in 5D, the spacetime symmetry group does not admit Weyl fermions as a faithful representation, since the Clifford algebra contains $\gamma^5$.  Instead, a 5D fermion is a 4-component Dirac fermion from the perspective of the 4D Lorentz subgroup. A dual picture of this is that a CFT (or AdS space) has spacetime symmetry $SO(4,2)$ (or cover $SU(2,2)$) rather than $SO(3,1)$ ($SL(2,\mathds{C})$). To obtain a left- or right-handed chiral fermion in the low-energy effective theory (as required by the structure of the SM), one chooses boundary conditions such that the opposite chirality has Dirichlet boundary conditions on both branes. For fermions, there are some puzzles in associating these 5D boundary conditions with 4D dual dynamics.  

On the UV brane, where it is understood that explicit conformal symmetry breaking is taking place in a dual description, the boundary condition (BC) is not so difficult to understand.  For our discussion, let us consider a $\left(\begin{smallmatrix}+\\-\end{smallmatrix}\right)$ boundary condition on the UV brane (where the RH fermion has the Dirichlet BC).  In this case, the dual picture contains a left-handed Weyl fermion in the fundamental sector, which is coupled to a spinor operator in the CFT sector~\cite{Contino:2004vy}.

On the IR brane, one can then choose either $\left(\begin{smallmatrix}+\\-\end{smallmatrix}\right)$ or $\left(\begin{smallmatrix}-\\+\end{smallmatrix}\right)$ boundary conditions. The former gives a massless LH Weyl fermion in the spectrum of modes, followed by a Kaluza-Klein tower.  The latter choice gives a massive tower of Dirac KK modes with no zero-mode.  Our focus is on the IR boundary condition, which is more difficult to interpret in a putative dual picture.

We would like to understand the IR fermionic boundary condition choices in terms of different options for  spontaneous breaking of the CFT through a pattern of vacuum expectation values (VEVs) of operators.  This is what the IR brane is supposed to represent. Such VEVs must somehow single out LH or RH operators to produce the spectrum we obtain from those boundary conditions. It is not obvious how to accomplish this in the 5D picture while respecting the 5D bulk spacetime symmetries.

The issue is laid bare by a calculation of anomalies associated with the classical phase rotation symmetry of a 5D fermion with $\left(\begin{smallmatrix}+\\-\end{smallmatrix}\right),\left(\begin{smallmatrix}+\\-\end{smallmatrix}\right)$  boundary conditions.  In such cases, it has long been known that the contributions to the anomaly are split across the two boundaries of the extra dimension:
\beq
\partial_M J^M = \frac{1}{4} \left[ \delta(z-z_0) + \delta (z-z_1) \right] {\mathcal Q}(z,x)
\eeq
where $J^M = \sqrt{g} \bar{\Psi} \gamma^M \Psi$, and ${\mathcal Q} = \frac{1}{16 \pi^2} F \cdot \tilde{F}$ corresponding to a background gauge field~\cite{Arkani-Hamed:2001uol}.  Note that in the cases where the gauge symmetry is broken on the branes, this represents a 't Hooft anomaly associated with residual global subgroups of the bulk gauge symmetry.

The full anomaly (integrated over the extra dimension and evaluated on the low-lying gauge field zero modes) is the same as the anomaly contribution in a 4D theory with a single Weyl fermion. It arises, however, from two \emph{geometrically isolated} contributions in the 5D theory.

The dual description of such a set-up is strange.  It seems that the fundamental sector (UV brane) and the CFT sector low-energy dynamics (IR brane) each contribute 1/2 of a Weyl fermion's amount to the anomaly. This appears inconsistent with the representation theory of the spacetime symmetries.  The minimum irreducible representations of the 4D Lorentz group, $SO(3,1)$, are single Weyl fermions (RH or LH).  The conformal group, on the other hand, $SO(4,2)$, has no chiral representations. Chirality (and the associated chiral anomaly) should be confined to the fundamental sector, and thus the UV brane in a dual geometry.

\begin{figure}[ht!]
\center{
\includegraphics[width=0.85\textwidth]{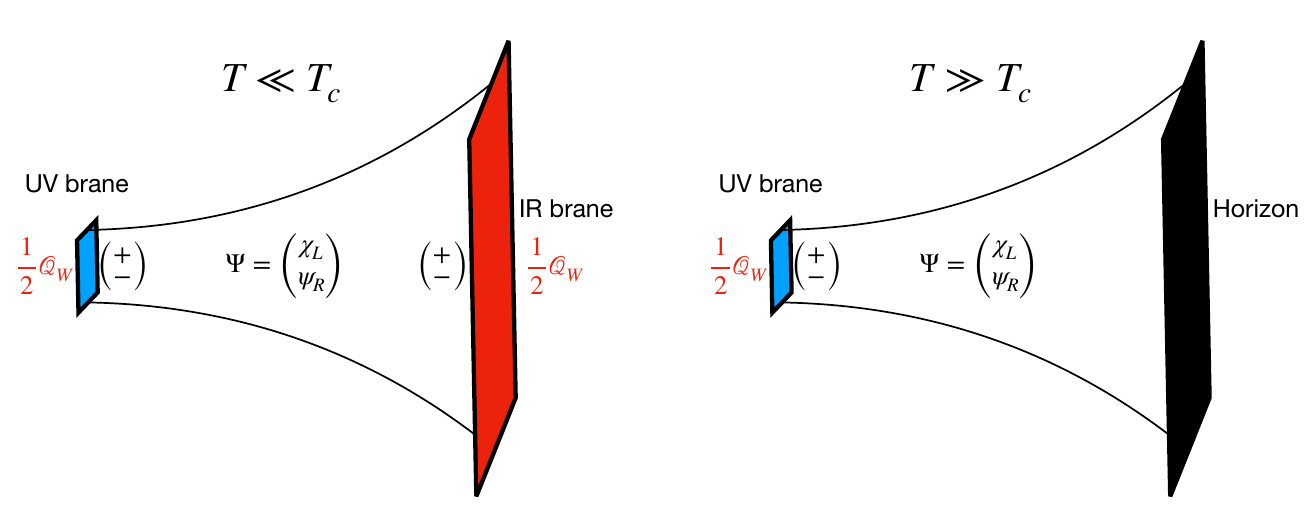}}
\caption{On the left is a geometry with both UV and IR branes corresponding to a ground (or low-temperature) state of an approximate CFT, in which the conformal symmetry is broken spontaneously.  On the right, a geometry with a horizon represents a high-temperature state where the conformal symmetry is unbroken.  $T_c$ is a critical temperature associated with the phase transition from the black brane phase to the IR brane phase.}
\label{fig:anomalycartoon}
\end{figure}

The problem of finding a sensible dual picture is compounded by the fact that different states - corresponding to the different IR truncations of the geometry - exhibit different anomaly structures, with the anomaly not always summing to a Weyl contribution in the low-energy 4D theory. In one case, we can consider a geometry that has an IR brane, and hence a localized IR anomaly contribution. In others, there is only a UV brane: at a high finite temperature, the dual geometry is AdS-Schwarzschild~\cite{Creminelli:2001th}, or some more general cosmology~\cite{Eroncel:2023uqf}, leaving no place to impose fermionic or gauge IR boundary conditions.  Alternatively, the system may be sitting in an early universe metastable state with some effective vacuum energy, driving an inflationary epoch in which the IR region of the bulk develops a horizon associated with de Sitter temperature. In either of these cases, there is no clear mechanism for imposing a chiral boundary condition.  When the IR brane is absent, the total anomaly appears to be 1/2 of a Weyl contribution. A cartoon of this is shown in figure~\ref{fig:anomalycartoon}, where a 5D fermion with $\left(\begin{smallmatrix}+\\-\end{smallmatrix}\right),\left(\begin{smallmatrix}+\\-\end{smallmatrix}\right)$ boundary conditions leads to an anomaly contribution that is split across the UV and IR branes.  

These various arguments point to the potential issues with having chiral IR brane boundary conditions while maintaining a consistent dual CFT description.  

The dual picture hints at a resolution, where the IR brane corresponds to spontaneous breaking of the CFT.  If a LH or RH degree of freedom is somehow decoupled by IR dynamics of the CFT, there should be an associated Wess-Zumino-Witten term from integrating it out~\cite{Wess:1971yu,Witten:1983tw}. Such terms, in a 5D description, arise from anomaly inflow from a bulk topological term such as a Chern-Simons contribution to the 5D Lagrangian~\cite{Callan:1984sa}. Adding such terms can remove the IR brane anomaly and relocate it to the UV brane.  In this case, there are no longer inconsistencies across different IR truncations of the same theory.

In this paper, we argue for the inclusion of 5D Chern-Simons or other topological anomaly-inflow terms when there are non-canceling fermionic contributions to anomalies on the IR brane. We calculate anomalies in various classes of hard-wall and horizon-ended geometries, show that the anomaly is not consistent across dual phases, and demonstrate that inflow terms can rectify the problem.  We further note that the UV content of the theory, specified by both UV brane fermionic boundary conditions and bulk inflow terms, then dictates the possible fermionic boundary conditions in the IR, whose anomaly must cancel against the contribution of the inflow terms.

The paper is organized as follows. In Section 2, we review anomalies in extra-dimensional theories. In Section 3, we introduce the different geometries of interest, including hard-wall Randall-Sundrum, 4D de Sitter slices, and AdS-Schwarzschild backgrounds. In Sections 4 and 5, we study the behavior of fermions and gauge fields in these geometries, with particular attention to their behavior near horizons and the implications for IR-localized anomalies. In Section 6, we show how topological terms can provide an anomaly flow that moves IR-localized anomalies to the UV.  We conclude in Section 7.

\section{Review of anomalies in extra dimensions}

The Adler-Bell-Jackiw $U(1)_A$ anomaly for a single Dirac fermion in 4D with its $U(1)_V$ symmetry gauged is encapsulated in the operator relation associated with the divergence of the axial current:
\beq
\partial_\mu j_5^\mu + 2 i m \bar{\psi}\gamma^5 \psi = {\mathcal Q}
\eeq
where ${\mathcal Q}$ is the anomaly, given by ${\mathcal Q} = \frac{1}{16 \pi^2} F \cdot \tilde{F}$~\cite{Adler:1969gk,Bell:1969ts}. Here $F$ is the field strength for a background potential $A_\mu$.  A massless Weyl fermion contributes similarly, though with the anomaly term being smaller by a factor of $1/2$.

In theories with compactified extra dimensions, there will be a discrete or continuous spectrum of modes for the gauge fields and fermions. These may involve massless gauge bosons and/or massless Weyl fermions.  The couplings of fermions to gauge fields, calculated on given modes, will have a non-trivial chiral structure.  The anomaly then involves a trace over the tower of LH and RH modes and their chiral charge:
\beq
\partial_\mu j_5^\mu +  i \sum_j m_j \bar{\psi_j}\gamma^5 \psi_j = \frac{1}{32 \pi^2}\Tr \left[ Q F \cdot \tilde{F} \right].
\eeq

The anomalies we are investigating in this work are anomalies in a 5D theory. In this case, there is a classically conserved 5D current, $J_M$, coupled to the gauge field $B_M$. Given a spectrum of fermion modes, we can calculate the anomaly of that current, which we expect will have some kind of extra-dimensional profile.

We know, however, that the bulk 5D theory is non-chiral, and we should therefore expect that contributions to the anomaly vanish everywhere except in regions where the 5D spacetime isometries are broken, admitting a localized chiral structure. The cancelations in the bulk must happen point by point and occur by summing up the anomaly contributions over the entire tower of modes.

We therefore expect that the divergence of the 5D current in a model with world-ending branes at positions $z_0$ and $z_1$ will take the form
\beq
\nabla_M J^M (x,z) = \chi(z) {\mathcal Q}(x,z)
\eeq
where $\chi(z)$ vanishes in the bulk, but may have non-vanishing contributions on branes. The specific form of $\chi(z)$ depends on the boundary conditions.  For $\left(\begin{smallmatrix}+\\-\end{smallmatrix}\right),\left(\begin{smallmatrix}+\\-\end{smallmatrix}\right)$ boundary conditions, $\chi(z) = \frac{1}{4} \left[ \delta (z-z_0)+ \delta (z-z_1) \right]$\cite{Arkani-Hamed:2001uol}. The magnitude of the coefficients in front of the $\delta$ functions is identical.  This feature is independent of the background geometry -- the anomaly is topological in nature~\cite{Hirayama:2003kk,Hong:2020xrh}. For different choices of boundary conditions, only the signs in front of the two delta functions will change.

Such an anomaly is not necessarily catastrophic for a 5D gauge theory, since boundary conditions might be assigned to the gauge fields that ensure the anomalous gauge transformation is only \emph{global} with respect to the 4D coordinates on the branes, setting ${\mathcal Q}$ to zero.  The brane-localized anomalies are then less harmful (but still physically relevant) 't Hooft anomalies. Another option is that there are mixed anomalies, and ${\mathcal Q}$ contains the gauge fields of other non-anomalous symmetries.

There is a correspondence between the 5D anomaly and the anomaly of the 4D low-energy effective theory. Under a 5D gauge transformation determined by the function $\beta(x,z)$, the total shift in the action under the transformation is given by
\beq
\delta S = \int d^5 x \beta(x,z) \chi(z) {\mathcal Q}(x,z).
\eeq
The effective 4D anomalous transformation is obtained by integrating over the extra-dimensional coordinate. For the case of the anomaly function above, the result is 
\beq
\delta S = \frac{1}{4} \int d^4 x \beta (x,z_0) Q(x,z_0) + \beta (x,z_1) Q(x,z_1) \ni \frac{1}{2} \int d^4 x \beta_0 (x) Q_0 (x)
\eeq
In the last part of this equation, we have focused on the portion of the anomaly that is evaluated on the zero mode gauge bosons (which have flat wave functions due to gauge invariance); this is the total 4D anomaly of a single massless Weyl fermion.

While 5D anomaly cancelation implies that the 4D anomaly vanishes, the converse is not necessarily true.  A fermion with $\left(\begin{smallmatrix}+\\-\end{smallmatrix}\right),\left(\begin{smallmatrix}-\\+\end{smallmatrix}\right)$ boundary conditions (for example) will lead to opposite-sign contributions to the anomaly on each brane, even though there are no zero-mode chiral fermions in the 4D effective theory.  The sum vanishes, so there is no 4D anomaly, but there \emph{is} an anomaly structure in the 5D theory, manifested on a non-trivial background of the KK-mode gauge fields.

Generally, for a single 5D fermion that has either $\left(\begin{smallmatrix}+\\-\end{smallmatrix}\right)$ or $\left(\begin{smallmatrix}-\\+\end{smallmatrix}\right)$ boundary conditions on each brane, the corresponding contribution to the anomaly will be associated with a function 
\beq
\chi(z) = \frac{1}{4} \left[ (\pm)_0 \delta (z-z_0) + (\pm)_1 \delta (z-z_1) \right]
\eeq
where $(\pm)_0 = + (-) 1$ when the RH (LH) component has Dirichlet boundary conditions on the brane at $z= z_0$, and $(\pm)_1 = + (-) 1$ when the RH (LH) component has Dirichlet boundary conditions on the brane at $z= z_1$.

Again, the anomaly coefficients are topological and independent of bulk curvature.  As we will argue, this leads to a puzzle when attempting to associate 5D geometries with states of a 4D theory when the 5D geometries develop horizons between geometrically separated contributions to the total anomaly.

\section{IR branes and Horizons}

For the purposes of this work, we will focus primarily on three geometries:
\begin{itemize}
    \item {\bf Randall-Sundrum I:} These are models in a static AdS or near-AdS background with two world-ending branes~\cite{Randall:1999ee,Randall:1999vf}.  They are dual to approximate conformal invariance spontaneously broken by VEVs of CFT operators~\cite{Rattazzi:2000hs,Arkani-Hamed:2000ijo}. These are presumed to be vacuum states with vanishing 4D effective cosmological constant.
    \item {\bf Metastable/Inflationary:} These are detuned quasi-stationary states of RS models where the UV brane tension is mismatched against the bulk cosmological constant so as to give an effective 4D cosmological constant.  This could arise dynamically as an excitation of a given RS theory through displacement of some scalar degree of freedom from its vacuum state.  The geometry could either be trapped in a metastable state, or correspond to some scalar degree of freedom slowly rolling down a flat potential.  In either case, the 4D slices are de Sitter, or approximately de Sitter.
    \item {\bf AdS-Schwarzschild:} An alternative solution to the Einstein equations, the AdS-Schwarzschild geometry features a bulk horizon, and is dual to exciting to a finite temperature state~\cite{Witten:1998zw}. This is relevant, e.g., in a radiation-dominated phase of big-bang cosmology.
\end{itemize}

All of these geometries can be taken to correspond, via AdS/CFT, to the same fundamental degrees of freedom and interactions, representing different states or phases of that same theory.  The point of our paper is to show that anomalies do not always match across these geometries without attention to canceling them in the spontaneously broken (UV and IR brane) phase.  These geometries are not intended to be exhaustive, but rather serve to sufficiently make the argument for the inclusion of either topological anomaly flow terms, or for new degrees of freedom that accomplish IR anomaly cancelation.

For the first two geometries, RSI and inflationary, the metrics we consider take the form:
\beq
ds_5^2 = A^2(z) \left(  ds_4^2 - dz^2 \right).
\label{eq:metric}
\eeq
For RSI two-brane geometries, the metric is static, and $ds_4^2$ is taken to be the 4D Minkowski metric.  For a pure $AdS$ background, with 5D cosmological constant given by $\Lambda = -6 k^2/\kappa^2$, the conformal factor $A(z) = 1/kz$, where $k$ is the AdS curvature.\footnote{Stabilization is required in order to give the radion a mass, and remove the flat direction associated with translation of the branes relative to each other.  The geometry then deviates from pure AdS due to backreaction effects of whatever mechanism is employed.  The details of this stabilization, however, are unimportant for the anomaly discussions.  We thus work with the simple original RSI model.}  Taking the UV brane to be located at $z = z_0$, and the IR brane to be located at $z_1$, the length of the extra dimension is $L_0 = k^{-1} \log (z_1/z_0)$, and the spectrum of various degrees of freedom includes a discrete KK tower beginning at $M_\text{KK} \approx z_1^{-1}$
which may begin with light modes depending on boundary conditions.

For metastable or inflating geometries, we take the 4D slices of the 5D geometry to be approximately de Sitter, in which case we have $ds_4^2 = \left( \frac{1}{H \eta}\right)^2 ds_\text{Mink}^2$, where $\eta$ is conformal time, $H$ is a constant Hubble rate, and $ds^2_\text{Mink}$ is the Minkowski line element.  The metric warp factor $A(z)$ that then solves the Einstein equations (presuming no matter content except for the bulk cosmological constant) is given by $A(z) = \frac{H}{k \sinh (H z)}$.  We obtain the usual 5D AdS geometry, $A = 1/kz$, in the limit $H \rightarrow 0$, or for $z \ll 1/H$.  The geometry is thus asymptotically AdS in the UV limit, $z\rightarrow 0$.  In the IR, the non-vanishing Hubble rate constitutes a low-energy deformation of the theory, and the geometry shuts off at a horizon as $z\rightarrow \infty$. Cutting off the geometry with a UV brane at $z= z_0 \ll 1/H$, the space is of finite length given by:
\beq
L_0 = \int_{z_0}^{\infty} \frac{H}{\sinh H z} dz \approx \log \frac{2}{H z_0}.
\eeq
The finite length again leads to a gap in the spectrum of fields propagating in the extra dimension.  Instead of a discretuum of KK-modes, however, one finds that a gapped continuum is the generic feature of the spectrum for fields propagating in the bulk~\cite{Cacciapaglia:2008ns,Guerrero:2019qqj}.  The gap itself has a dual description in terms of a finite dS temperature, $T = \frac{H}{2 \pi}$, though its effects appear already at the classical level in the 5D theory.

There are other time-dependent geometries that cut off the fifth dimension at some finite length, with a horizon ending the geometry.  These include 5D cosmologies with more generic FLRW 4D slices, such as those corresponding to radiation domination. However, these do not necessarily have a 4D cosmology which ``factorizes'' from the 5D conformal factor as in Eq.~\ref{eq:metric}, exhibiting a more complicated $z$ and $t$ dependence, or a moving UV brane. 

The geometries associated with a radiation-dominated cosmology, however, have a close cousin, which is the AdS-Schwarzschild geometry, or AdS with a bulk black hole~\cite{Gubser:1999vj,Gursoy:2008za}. We can consider the geometry in Euclidean signature, matching a compactified time radius to the Hawking temperature (corresponding to thermal equilibrium between the black hole and the bulk).
The bulk metric for the AdS-Schwarzschild geometry is
\begin{equation}
ds^2 = A^2(z) \left( h(z) dt^2 + d\vec{x}^2 + h^{-1} (z) dz^2 \right).
\end{equation}
The solution in the case of a pure bulk cosmological constant is 
\begin{equation}
A(z)  = \frac{1}{k z}, ~~~~~ h(z) = 1 - \left( \frac{z}{z_h} \right)^4
\end{equation}
where $z_h$ is the horizon location.

For our study of the behavior of bulk field solutions, which are relevant for the anomaly calculation, it is better to express the metric in ``tortoise" coordinates, where the 00 and 55 components of the metric are the same:
\begin{equation}
\label{eq:AdSS_metric}
ds^2 = A^2(z) \left( h(z) dt^2 + d\vec{x}^2 + h (z) d\tilde{z}^2 \right).
\end{equation}
In this equation, it is implied that $z$ is a function of $\tilde{z}$, with the mapping between coordinate systems obtained by inverting the following relation:
\beq
\tilde{z} = \frac{1}{2} z_h \left[ \tan^{-1} \left( \frac{z}{z_h} \right) + \frac{1}{2}\log \frac{1+ z/z_h}{1-z/z_h} \right] \xrightarrow{z \rightarrow z_h}  \frac{z_h}{4} \log \frac{2}{1-\frac{z}{z_h}}.
\eeq
In these coordinates, the AdS black hole is at $\tilde{z}\rightarrow \infty$.

\section{Fermion Wave Functions}

For the purpose of calculating anomalies, we require the fermion wave functions that solve the Dirac equation in the geometries we just reviewed.  The 5D Dirac action is given by
\beq
S_\Psi = \int d^5x \sqrt{g} \bar{\Psi} \left( i \overleftrightarrow{\slashed{D}} + m \right) \Psi.
\eeq
The $\overleftrightarrow{\slashed{D}}$ is an antisymmetric difference operator, $\overleftrightarrow{\slashed{D}} = \frac{1}{2} \left( \overrightarrow{\slashed{D}}-\overleftarrow{\slashed{D}} \right)$, introduced here to ensure hermiticity in the presence of boundaries of the 5-dimensional space.  We will first focus on metrics of the form in Eq.~\ref{eq:metric} before studying the case of the AdS-Schwarzschild geometry in~\ref{eq:AdSS_metric} separately.

\subsection{de Sitter slices}

The 5D Dirac equation obtained by variation of the action above in a warped background is given by:
\beq
{\mathcal D}_{4} \Psi + \gamma^5 \partial_z \Psi + \left( 2 \frac{A'}{A} \gamma^5+ A k c \right) \Psi = 0,
\eeq
where we take ${\mathcal D}_{4}$ to be the 4D Dirac operator associated with inflating 4D slices, and $c$ is the Dirac mass in units of the 5D curvature, $k$.

The equation of motion is simplified by a rescaling of the fermion fields:  $\Psi = \frac{\psi}{A^2}$.  In this case, the equation for $\psi$ is:
\beq
{\mathcal D}_{4} \psi + \gamma^5 \partial_z \psi +  A k c \psi = 0.
\eeq
As $A(z)$ is a monotonically decreasing function for AdS or geometric shut-offs, this tends towards the massless 5D Dirac equation in the limit $z\rightarrow \infty$.  

To solve the 5D Dirac equation, we expand the 5D spinor in terms of 4D Weyl spinors:
\beq
\psi_\mu = \left( \begin{array}{c} g_\mu(z) \chi_\mu (x) \\ f_\mu(z) \bar{\psi}_\mu(x) \end{array} \right).
\eeq
 $\chi_\mu$ and $\bar{\psi}_\mu$ are eigenmode pairs of the 4D Dirac equation with eigenvalue $\mu$. The functions $f_\mu(z)$ and $g_\mu(z)$ are the corresponding 5D wave functions, which satisfy:
\begin{align}
f' &+ \mu g - A k c f=0 \nonumber \\
g' &- \mu f + A k c g=0.
\end{align}
This can be written as a matrix equation, 
\begin{equation}
\label{eq:wavefunctions}
    \begin{pmatrix} f \\ g \end{pmatrix}' = \hat{M} \cdot \begin{pmatrix} f \\ g \end{pmatrix},~~~\mathrm{with}~~~ \mathbf{M} = \begin{pmatrix}  A k c & - \mu \\ \mu & -A k c \end{pmatrix}.
\end{equation}
Such equations can at least be formally integrated, with the result being given by a path-ordered exponential integral:
\beq
\begin{pmatrix} f(z) \\ g(z) \end{pmatrix} = \mathbf{P} \exp \left[ \int_{z_0}^z \mathbf{M}(\tilde{z}) d\tilde{z} \right] \cdot \begin{pmatrix} f(z_0) \\ g(z_0) \end{pmatrix}.
\eeq

\noindent
{\bf Zero Mode Fermions}

For $\mu = 0$, the corresponding wave function is obtained by integrating Eq.~\ref{eq:wavefunctions} out from the UV brane at $z = z_0$.  As $\hat{M}$ is self-commuting at all $z$ for $\mu = 0$, this has a simple analytic expression:
\beq
f_0(z) \propto  e^{+c k L(z)}, ~~ g_0(z) \propto  e^{-c k L(z)}~~~~~\mathrm{with}~~~L(z) \equiv \int_{z_0}^{z} A(\tilde{z}) d\tilde z.
\eeq
$L(z)$ is just the proper distance into the bulk from the UV brane to point $z$.  

The UV brane boundary conditions admit left- (right-) handed zero-mode excitations if $\bar{\psi}_{z=z_0} = 0$ ($\chi_{z=z_0} = 0$).  However, the UV boundary condition is not sufficient to determine whether a zero mode exists. We also require normalizability (or appropriate boundary conditions if there is a hard-wall IR brane).

To check normalizability, we consider the kinetic term of a massless eigenmode in the 4D effective theory.  The 5D Dirac term contains:
\beq
\int d^5x \sqrt{g} \bar{\Psi} \gamma^\mu e_\mu^M \partial_M \Psi \supset \int d^5x A^4 \sqrt{g_4} \bar{\Psi} \gamma^\mu (e_4)_\mu^\nu \partial_\nu \Psi,
\eeq
where, in the second expression, we have factored out the metric $A(z)$ function as it appears in the 5D metric and the f\"unfbein.  For a $\bar{\psi}_0(x)$ zero mode with wave function $f_0$, the effective 4D kinetic term is given by:
\beq
\int_{z_0}^{\infty} dz f_0^2  \int d^4x \sqrt{g_4} \bar{\psi}_0 \gamma^\mu (e_4)_\mu^\nu \partial_\nu \psi_0.
\eeq 
The rescaling performed above has canceled out the overall $A^4$.
For the mode to be normalizable, we require $\mathcal{N}^2=\int_{z_0}^{\infty} dz f_0^2 = \int_{z_0}^{\infty} dz e^{2c k L(z)}$ to be finite.  If $L(z)$ approaches a constant for large $z$, then the integral does not converge, and there is no normalizable zero mode.  If the geometry is cut off at large $z = z_1$ with a hard IR brane, a zero mode is allowed, assuming one imposes suitable IR-brane boundary conditions.  Note that a similar argument can be applied to a potential $\chi_0(x)$ zero mode.  

The above is an interesting result:  in infinite RS, where $A(z) = \frac{1}{z}$ and $L(z) = \log z/z_0$, it is well known that there is a potential $\bar{\psi}_0(x)$ zero mode for $c < -1/2$ (or a $\chi_0(x)$ zero mode for $c>1/2$).  However, for soft-wall shutoffs like the one we study, where $L(z) \rightarrow L_0$, such modes are not normalizable.  It is somewhat unusual that truncation of a geometry renders modes non-normalizable.

\noindent
    {\bf The Dirac Spectrum}

To determine the spectrum of modes with non-vanishing Dirac eigenvalues, it is helpful to put the equations into Schr\"odinger form.  This is readily accomplished by taking the derivative of the matrix equation,~\ref{eq:wavefunctions}: 
\begin{align}
&f'' + \left[ \mu^2 -A' k c - A^2 k^2 c^2 \right] f = 0 \nonumber \\
&g'' + \left[ \mu^2 +A' k c - A^2 k^2 c^2 \right] g = 0
\end{align}
These are Schr\"odinger type equations with  ``potential" terms $V_\text{eff} = A^2 k^2 c^2 \pm A' k c$.  The behavior of this effective potential as $z\rightarrow \infty$ contains information about whether the modes will form a continuum or discretuum, and whether there will be a gap.  For geometric shutoffs of the geometry, $L(z) \rightarrow \text{constant}$, so any derivative of $L$ must trend towards zero as $z\rightarrow \infty$.  Since $A = L'$, and $A' = L''$, we find that no such shutoff of the geometry can gap the fermions.  The spectrum of the fermions in the geometry that ends with a horizon is an ungapped continuum with a potential UV-localized zero mode in non-compact geometries with $L \rightarrow \infty$.

We can study the behavior of solutions by analyzing Eq.~\ref{eq:wavefunctions}.  The path-ordered exponential integral does not generally take on a simple analytic form, and even when it does, that form is not always illuminating.  However, we can use the fact that the integral approximately separates into regions to determine the asymptotic behavior. This separation is due to the fact that $A(z)$ is a monotonically decreasing function.  Noting that the matrix $\hat{M}$ can be expanded in terms of Pauli matrices as 
\beq
\mathbf{M}(z) = \left( \begin{array}{cc} c k A(z) & -\mu \\ \mu & -c k A(z)  \end{array} \right) = c k A(z) \hat{\sigma}_3 - i \mu \hat{\sigma}_2,
\eeq
we see that there will always be some $z_*$ at which the constant $\mu$ and the decreasing $A$ cross.  The solution above can then be approximated by a multiple of two matrices formed by the integration over the regions $z>z_*$ and $z < z_*$.  In each of these regions, the matrix is proportional to either of $\hat{\sigma}_{2,3}$, which are self-commuting for $z$ in those regions.  In each region, the path integral is simple, with only the cross-over behavior posing difficulties:
\beq
\label{eq:softwallsols}
\begin{pmatrix} f(z) \\ g(z) \end{pmatrix}\approx \left\{ \begin{array}{ll} 
\exp \left[ c k L(z) \hat{\sigma}_3 \right] \begin{pmatrix} f(z_0) \\ g(z_0) \end{pmatrix} & z \ll z_* \\
\begin{pmatrix}
    \cos \mu z & - \sin \mu z \\ 
    \sin \mu z &  \cos \mu z
\end{pmatrix} \cdot\mathbf{\Sigma}\cdot \exp \left[ c k L(z_*) \hat{\sigma}_3 \right]  \begin{pmatrix} f(z_0) \\ g(z_0) \end{pmatrix} & z \gg z_* 
 \end{array} \right.
\eeq
$\mathbf{\Sigma}$ is a constant matrix that encodes details of the transition between the two asymptotic regions of the integral.

We are generally after the behavior of solutions in the large-$z$ region.  We are looking for any sign that a chiral boundary condition can be imposed where an anomaly that lives on the IR brane in one phase can survive as anomaly information localized on a horizon at $z \rightarrow \infty$ in another.  The answer is clear in Eq.~\ref{eq:softwallsols}.  The near-horizon behavior of the fermion wave functions is the same as in an infinite 5D flat bulk.  There is no place to enforce a chiral boundary condition, and thus no way for the fermions to contribute to a chiral anomaly in the near-horizon region.

\subsection{AdS-Schwarzschild}

We now consider the case in which fermions are in the AdS-Schwarzschild geometry. The 5D Dirac equation for a fermion, $\Psi$, in the tortoise coordinates employed in Eq.~\ref{eq:AdSS_metric} is most conveniently written by first performing a rescaling $\Psi = A^{-2} h^{-1/4} \psi$:
\begin{equation}
\left[ \gamma^0 \partial_t + \gamma^5 \partial_{z} + \sqrt{h} \left( \vec{\gamma} \cdot \vec{\partial} + c k A \right) \right] \psi = 0.
\end{equation}
Note that we are using $z$ in place of $\tilde{z}$ for simplicity of notation.  In the near-horizon limit ($z \rightarrow \infty$) we have $h \rightarrow 0$, and $A \rightarrow \frac{1}{k z_h}$ (here, $z_h$ is the horizon location in the original AdS-Schwarzschild coordinates, so $A$ goes to a constant).  
We now consider a mode with some fixed three-momentum $\vec{k}$. In the approach to the horizon, we can leave out the constant terms multiplied by $\sqrt{h}$, and the equation is simply
\begin{equation}
\left( \gamma^0 \partial_t + \gamma^5 \partial_{z} \right) \psi = 0, ~~~\mathrm{or}~~~\begin{pmatrix} f \\ g \end{pmatrix}' = -i \omega \hat{\sigma}_2 \begin{pmatrix} f \\ g \end{pmatrix}
\end{equation}
where we have expanded on energy eigenstates in the second equation.  This is of the same basic form as the near-horizon wave functions in the previous geometry, with wave functions oscillating indefinitely into the horizon, mixing LH and RH modes.  We conclude again that there is no natural chiral boundary condition that one can impose such that the horizon can sustain anomaly information.

\section{5D Gauge Fields}

A 5D gauge symmetry in an AdS geometry is dual to the existence of a conserved global current, $J^\mu$, in a dual CFT.  The boundary conditions chosen for the gauge field on a UV cutoff brane correspond to whether the current couples to a fundamental gauge field.  Dirichlet boundary conditions for the 4-vector potential, $B_\mu$, correspond to keeping the symmetry a global symmetry, and vice versa for Neumann boundary conditions.

If there is an IR brane, Dirichlet conditions there for $B_\mu$ correspond to the spontaneous breakdown of the global symmetry to which $J^\mu$ is associated.  If that global symmetry was gauged, the gauge theory is in a Higgs phase. If it was not gauged, there is a $B_5$ zero mode that nonlinearly realizes the global symmetry, which is spontaneously broken.

Before considering the presence of possible anomalies, we would like to better understand what the correspondence entails for geometries that end on a horizon rather than a hard wall.  

A pure gauge theory in 5D has an action given by
\begin{equation}
S=\int d^5x \sqrt{g}\left[-\frac{1}{4}
B_{MN} B_{PQ} g^{MP}g^{NQ}\right].
\end{equation}
We consider this action for both the de Sitter-slice geometry and the AdS-Schwarzschild geometry.

\subsection{de Sitter slices}

On the metric in Eq.~\ref{eq:metric}, this is:
\begin{equation}
S=\int d^4x \int_{z_0}^{\infty}\ dz A(z) \left[ -\frac{1}{4} {B_{\mu\nu}}^2
+\frac{1}{2} {B_{\mu 5}}^2 \right],
\end{equation}
where 4D metric contractions are performed using $\eta_{\mu\nu}$, and all 5-indices are lowered.
The second term contains kinetic mixing terms between the $B_5$ and the $B_\mu$:
\begin{equation}
\int d^4x \int_{z_0}^{\infty} dz A(z) B_5 \partial_\mu \partial_5 B^\mu =
-\int d^4x\int_{z_0}^{\infty}  dz\partial_\mu B^\mu \partial_5 \left( A(z) B_5
\right) - \int d^4 x A(z_0) B_5 \partial_\mu B^\mu
\end{equation}
where we have integrated by parts. Independent gauge-fixing terms in the bulk and on the boundary that remove this mixing are:
\begin{align}
S_{gf}=&-\int d^4x \int_{z_0}^{\infty}dz \frac{1}{2\xi} A(z) \left[
\partial_\mu B^\mu - \frac{\xi}{A(z)} \partial_5 \left(A(z) B_5
\right)\right]^2 \nonumber \\
&-\int d^4x \frac{1}{2\xi_0} A(z_0) \left[
\partial_\mu B^\mu - \xi_0 B_5
\right]^2.
\end{align}

The bulk equations of motion for the $B_\mu = b(z) \epsilon_\mu(p) e^{ipx}$ and $B_5 = a(z) e^{ipx}$ are:
\begin{align}
    \frac{1}{A} \partial_5 (A(z) \partial_5 b)+\mu^2 b &= 0 \nonumber \\
    \xi \partial_5 \left[ \frac{1}{A(z)} \partial_5 \left( A(z) a \right) \right] + \mu^2 a &=  0
    \label{eq:bulkgauge}
\end{align}
where $p^2 = \mu^2$ are the eigenvalues of the de Sitter box operator acting on the gauge fields.  These are obtained as usual via the variational principle, which also gives terms on the boundaries of the space:
\begin{align}
    \left. \delta B^\mu \partial_5 (B_\mu) \right|_\partial &= 0 \nonumber \\
    \left.  \left[ \xi \partial_5 \left( A(z) B_5 \right) + \xi_0 A B_5 \right]\delta B_5 \right|_\partial &= 0.
    \label{eq:boundgauge}
\end{align}
We now study zero modes and KK-modes in this geometry.

\noindent
{\bf Zero Modes}

There is a 4D effective gauge theory if there is a zero mode solution to the 5D $B_\mu$ bulk equation of motion and boundary conditions.  Such solutions to Eqs. (\ref{eq:bulkgauge}) and (\ref{eq:boundgauge}) correspond to constant wave functions.  In the AdS/CFT dictionary, this choice of boundary conditions corresponds to gauging a global symmetry of the CFT.

Integrating the zero mode wave function over the extra dimension yields an effective 4D gauge coupling:  $g^{-1}_4 = g^{-1}_5 \int dz A = g^{-1}_5 L_0$.  If $L_0$ diverges, corresponding to an infinite-size extra dimension, the effective gauge coupling goes to zero.  In other words, a potential zero mode gauge boson is not a normalizable state if the extra dimension is of infinite size.  On the other hand, if there is a shutoff of the geometry, as with a horizon, the integral is finite, and there is a prediction for the IR value of the effective gauge coupling.

In the AdS/CFT picture, the 4D gauge coupling going to zero is effectively the statement that at very long distances/low energies, the gauge coupling flows to zero when there are charged and ungapped degrees of freedom in the CFT -- the theory is IR free.  We can conclude that if the extra dimension is of infinite size, there can be no zero mode.  However, it is not correct to surmise that there is no gauge symmetry in the 4D dual theory in this case.  Any finite energy process corresponds roughly to a depth into the bulk.  I.e., a brane-to-brane scattering process with 4-momentum transfer $Q$ will, in a 5D calculation, be centered at position $z \sim Q^{-1}$. At such energies, the theory is fully 5-dimensional, or in the CFT language, the states of the CFT cannot be integrated out to give a perturbative effective 4D gauge theory.  In fact, the modes span an ungapped continuum when the length of the geometry is infinite.

The take-away lesson when we calculate anomalies due to fermions is that we should think of all geometries with a UV boundary condition $\partial_5 B_\mu = 0$ as ones where the UV anomaly must be fully canceled.

The above discussion has focused on the case when the global CFT current is gauged.  We can also consider the case where we take Dirichlet conditions for $B_\mu$.  In this case, there may be a massless scalar degree of freedom associated with $B_5$.\footnote{Massive elements of this tower can be gauged away -- they are the eaten modes of the massive 4D vectors.  Indeed, their spectrum is gauge-dependent as one sees in Eq.~\ref{eq:bulkgauge}}. From the equations of motion and boundary conditions, we see that the wave function of such a mode is given by $a(z) = A^{-1} (z)$.  From the kinetic term for the $B_5$, we see that such a mode is normalizable only if the integral $\int_{z_0}^{\infty} A^{-1} (z) dz$ is convergent.  However, if $L(z)$ approaches a constant, $A(z)$ needs to decrease faster than $1/z$, and the integral over the $B_5$ wave function can never be convergent.  This makes perfect sense in the context of the AdS/CFT correspondence.  A bulk horizon corresponds to a finite temperature, which is an explicit, not spontaneous, breaking of conformal invariance. There are thus no VEVs of operators, and so the global symmetry of the CFT cannot be broken.  There is thus no Goldstone boson.

\noindent
{\bf Kaluza-Klein Spectrum:}
  
The massive $B_\mu$ mode spectrum can be quantified by putting the equation of motion in Schr\"odinger form.  This is accomplished with the rescaling $\tilde{b} = A^{-1/2} b$:
\begin{equation}
    - \tilde{b}'' + \left[ \frac{1}{2} \frac{A''}{A} - \frac{1}{4} \left( \frac{A'}{A} \right)^2 \right] \tilde{b} = \mu^2 \tilde{b}.
\end{equation}
Unlike the fermion spectrum, the gauge fields can be gapped by a geometric shutoff.
Consider an $A(z)$ which decays exponentially in the asymptotic region near the horizon:  $A(z) \propto e^{-\mu_* z}$, with $\mu_*$ being some scale.  For such geometries, $V_\mathrm{eff} \rightarrow \frac{1}{4} \mu_*^2$.  For power-law decay, $A(z) \propto 1/z^\alpha$, the effective potential asymptotes to zero.

The geometry associated with dS slices corresponds to $A = \frac{H}{k \sinh Hz} \rightarrow \frac{2 H}{k} e^{- H z}$, so the effective potential asymptotes to $V_\mathrm{eff} \rightarrow \frac{1}{4} H^2$.  The solutions thus span a continuum above $\mu = \frac{1}{2} H$.  The solutions are therefore dS plane waves in the near-horizon limit.

\subsection{AdS-Schwarzschild}

For the AdS-Schwarzschild geometry, the spectrum is different.  The equations of motion for the transverse and longitudinal modes are
\begin{align}
    B_T'' &+\frac{A'}{A} B_T' + \Omega B_T = 0 \nonumber \\
    B_L'' &+\left( \frac{A'}{A} - \frac{\Omega'}{\Omega} \right) B_L' + \Omega B_L = 0,
\end{align}
where we have defined $\Omega = \omega^2 - h(z) \vec{k}^2$.

In the approach to the horizon, we have $A \rightarrow \frac{1}{k z_h}$ as $h \rightarrow 0$, so $A'/A$ vanishes.  Additionally, the three-momentum contribution to $\Omega$ is damped by the horizon function, $h$, and only the energy term, $\omega^2$, survives.  The near-horizon behavior of the equations is thus:
\begin{equation}
    B_{T,L}'' + \omega^2 B_{T,L} = 0.
\end{equation}
The gauge field spectrum is thus determined by the allowed frequencies, $\omega$.  In a thermal description, time is compactified on a circle of radius $T^{-1}$, and the spectrum consists of the usual Matsubara frequencies, $\omega_n = 2 \pi n T$.  Importantly, the modes oscillate all the way into the horizon.  

We note that there is no admissible/normalizable scalar zero mode.  In the absence of an IR cutoff brane, a $B_5$ or extra-dimensional Wilson line degree of freedom can be gauged away.  

\subsection{Gauge Theories on the Horizon}

The conclusion that we draw from the previous subsections is that in the geometries with horizons that we have studied, the 5D gauge theory persists unbroken into the horizon.  Any anomaly in this region of the geometry would violate the 5D gauge symmetry, and is thus forbidden by consistency.

\section{Anomaly Flow}

\subsection{Abelian Chern-Simons Terms}
Chern-Simons (CS) terms in 5D have phenomenological implications for 4D effective theories~\cite{Hill:2006ei,Collie:2008vc}.  In this work, we claim that such terms (or additional fermion content) are necessary for a sensible dual in some constructions.  We thus review some basic aspects of their inclusion.

For a $U(1)$ gauge theory in 5D, terms of the form
\begin{equation}
    \begin{aligned}
    {\mathcal L}_{\mathrm{CS}} &= c \, \epsilon^{ABCDE} B_A \partial_B B_C \partial_D B_E 
       = \frac{c}{4} \, \epsilon^{ABCDE} B_A F_{BC} F_{DE}
     \end{aligned}
\end{equation}
can be added to the Lagrangian.  Under a gauge transformation
\begin{equation}
    B_A \;\to\; B_A + \partial_A \theta,
\end{equation}
the action has a non-trivial transformation on the boundaries of the space \cite{Hill:2006ei}.  For example, in an RS setup with end-of-world branes at $z=z_0$ and $z=z_1$,
\begin{equation}
    S_{CS}\;\to\; S_{CS} + \frac{c}{4} \int d^4x \, \theta (z_0) \,\epsilon^{BCDE} \, F_{BC} F_{DE}(z_0)- \frac{c}{4} \int d^4x \,  \theta (z_1) \, \epsilon^{BCDE} \, F_{BC} F_{DE}(z_1) 
\end{equation}
\[\;\to\; S_{CS} + c \int d^4x \, \int_{z_0}^{z_1} dz \, \theta (z) \left[\delta(z-z_0) -\delta(z-z_1)\right]\, F_{BC} \tilde{F}_{DE}\]
where $\tilde{F}^{\mu\nu} = \tfrac{1}{2} \, \epsilon^{\mu\nu\rho\sigma} F_{\rho\sigma}$.

The Chern-Simons term thus generates an ``anomaly" at the classical level in the extra-dimensional theory given by
\begin{equation}
\partial_M J^M = \frac{c}{2} \left[ \delta(z-z_0) - \delta (z-z_1) \right] F \cdot \tilde{F}.
\end{equation}
The CS gauge variation does not give any contribution to the effective 4D anomaly, as the terms on the two branes are of opposite sign.  Evaluated on (flat) zero-mode gauge bosons, the contributions from the two branes cancel.

The CS term can, however, be utilized as an anomaly-flow mechanism.  With an appropriate choice of the CS coefficient, $c$, it can remove a fermionic IR brane contribution to the anomaly, moving it to the UV brane.  

Consider, for example, the contribution to the anomaly of a LH Weyl zero mode, obtained by taking $\begin{pmatrix} \pm \end{pmatrix}, \begin{pmatrix} \pm \end{pmatrix}$ boundary conditions:
\begin{equation}
\partial_M J^M_\mathrm{LH} = \frac{1}{64 \pi^2 } \left[ \delta(z-z_0) + \delta (z-z_1) \right] F \cdot \tilde{F}.
\end{equation}
This theory, based on the arguments above, will not have an anomaly consistent across phases. A horizon could envelop the contribution at $z_1$, leaving 
\begin{equation}
\partial_M J^M_\mathrm{LH} = \frac{1}{64 \pi^2 } \delta(z-z_0) F \cdot \tilde{F}.
\end{equation}
which is $1/2$ the contribution of a single Weyl fermion.

However, one can choose a Chern-Simons term coefficient such that it cancels the anomaly in the IR and adds it to the UV-localized anomaly.  In the example above, taking
$c = \frac{1}{32 \pi^2}$ gives
\begin{equation}
\partial_M J^M_\mathrm{LH+CS} = \frac{1}{32 \pi^2 } \delta(z-z_0) F \cdot \tilde{F}.
\end{equation}
With the addition of the CS term, the anomaly structure can now be constant across other states of the same dual CFT.  Moreover, the IR boundary-condition information is not completely lost when obscured by a horizon, but it is stored in the CS term.  That is, if one starts in the phase with a horizon, but then considers a transition to the phase with an IR brane, the CS term, along with the UV anomaly information, dictates which IR brane fermionic boundary conditions are admissible.

This last statement is worth emphasizing with an example.  Consider studying a theory that begins in a 4D dS slice inflating phase, and has a Dirichlet boundary condition for a 5D gauge field that couples to a fermion with either $\begin{pmatrix} \pm \end{pmatrix}$ (LH) or $\begin{pmatrix} \mp \end{pmatrix}$ (RH) boundary conditions on the UV brane.  In this phase, an observer can recognize the UV boundary condition for a fermion and can detect the presence and coefficient of a CS term.  If the observer sees a LH or RH Weyl contribution to a chiral anomaly on the UV brane,\footnote{Note that it would be a 't Hooft anomaly, as the gauge symmetry is restricted on the UV brane} constructing the anomaly coefficient as the sum of the consistent anomaly contribution and the CS coefficient, anomaly cancelation on an eventual IR brane/conformally broken phase  \emph{requires} a boundary condition for the fermion that gives a RH or LH Weyl fermion, respectively.  If the same observer detects a cancelation between the CS and fermionic contribution, the observer would find that the only admissible boundary condition on an IR brane would be one that does not admit a zero mode.

In a more complicated theory with more fermionic degrees of freedom, the restrictions on boundary conditions would be slightly less predictive for the spectrum, with the only requirement being net cancelation of all IR brane anomalies.

\subsection{Non-abelian Chern-Simons Terms}

Let us consider a non-abelian bulk gauge symmetry, $G$, and let us say that this symmetry is broken by boundary conditions on the IR brane to the subgroup $H_1$.

A bulk Chern-Simons term that is invariant under $G$ up to the usual boundary variation contributes (classically) to an anomaly:
\begin{equation}
\label{eq:nonabelianCS}
Q^a_\text{CS} = \frac{c}{2} \left[ \delta(z-z_0) - \delta (z-z_1) \right]\epsilon^{\mu\nu\rho\sigma} \Tr \left[ T^a \{ F_{\mu\nu}, F_{\rho\sigma} \} \right]
\end{equation}
The trace factor is over the group generators in the fundamental representation, and is proportional to the completely symmetric tensor, $d_G^{abc}$, defined by $\{T^a,T^b \} = \frac{1}{d} \delta^{ab} + d_G^{abc} T^c$.  Field strength terms aligned with the broken generators would be vanishing, though one can still interpret Eq.~\ref{eq:nonabelianCS} as an IR brane 't Hooft anomaly along those broken directions.

A straightforward generalization of the discussion above allows one to use the non-abelian CS term to flow an anomaly proportional to $d_G^{abc}$ from the IR into the UV.  This, however, does not cover all possible chiral structures. It would require that, for a bulk fermionic multiplet carrying $G$-charge, IR brane boundary conditions respect the original group $G$.  In principle, the gauge-breaking structure $G \rightarrow H_1$ permits other assignments. In such cases, the CS term cannot accomplish the required anomaly flow.

\subsection{Other anomaly flows}

Let us go back to the discussion above, with gauge group $G$ broken to subgroup $H_1$ on the IR brane.  Since the gauge group on the brane is smaller than $G$, the contribution to the anomaly from bulk fermionic fields carrying $G$-charge may not be proportional to $d_G^{abc}$.  That is, the consistent anomaly may give a contribution on the IR brane given by
\beq
Q^a_\text{IR} = \lambda \delta (z-z_1) \epsilon^{\mu\nu\rho\sigma} M^{abc} F_{\mu\nu}^b F_{\rho\sigma}^c.
\eeq
Here $M^{abc} \ne d^{abc}$ for general cases of boundary conditions that respect only $H_1$ transformations.  In such cases, the Chern-Simons term cannot remove any anomaly from the IR brane.

As a simple example, take the bulk gauge group $G = SU(2)$, and the IR brane localized gauge symmetry as the $U(1)$ diagonal subgroup.  There can not be a Chern-Simons anomaly, since $d^{abc}_{SU(2)} = 0$. Take the fermion content to be a single bulk fermion $SU(2)$ doublet.  Under the $U(1)$ subgroup, one component of the doublet, $\Psi_+$, carries charge $+1$, while the other, $\Psi_-$, carries charge $-1$.  Since $SU(2)$ is broken on the brane, the boundary conditions only need to satisfy the local $U(1)$ gauge symmetry.  One can thus choose $\mp$ boundary conditions for $\Psi_+$, and $\pm$ boundary conditions for $\Psi_-$.  This results in a gauge anomaly for the $U(1)$ given by $Q^3_\text{IR} =  \frac{1}{32 \pi^2} \delta (z-z_1)  \epsilon^{\mu\nu\rho\sigma} B_{\mu\nu} B_{\rho\sigma}$, where $B_{\mu\nu}$ is the field strength associated with the unbroken $U(1)$ subgroup. The anomaly in this case cannot flow from the IR into the UV with a CS term.

One can, however, easily flow the anomaly without introducing any new light degrees of freedom. Adding a second bulk fermion $SU(2)$ doublet with the opposite boundary condition assignments in the IR cancels the $U(1)$ anomaly.  The spectrum of light fermions is unchanged with UV brane boundary conditions that create the opposite anomaly contribution in the UV.  One can then consider integrating out this bulk fermion, with its contribution to the anomaly remaining as an effective topological term.  This type of anomaly flow may be captured by the more general $\eta$-invariant~\cite{Witten:2019bou}, of which CS terms form a particular sub-class with a local Lagrangian interpretation.

\section{Conclusions}

We have studied the geography of chiral anomalies in approximately AdS 5D theories where the geometry ends in the IR with either an  IR brane or a horizon.  We have described a tension in which horizons of the class we studied cannot manifest anomalies through either fermion loops or topological terms.  The horizon is not a boundary of the space. On the other hand, fermionic boundary conditions on an end-of-world IR brane are known to contribute to such anomalies.

In order to realize these distinct geometries via the AdS/CFT correspondence as different states of the same theory, we have concluded that all anomalies on a putative IR brane must be canceled.  This necessitates either new fermionic degrees of freedom or an anomaly-flow mechanism that shifts any anomalies resulting from IR brane fermionic boundary conditions to the UV boundary.  While our discussion has focused on chiral anomalies, we anticipate it should be generalizable to other non-perturbative and discrete anomalies.

Conversely, we argue that one can unambiguously read off the anomaly content of a theory from the summed contributions of bulk fermions (given their UV brane boundary conditions) and anomaly inflow terms.  This UV data places restrictions beyond simple gauge invariance for allowed fermionic boundary conditions on an IR brane phase of the theory.

\section*{Acknowledgements}
This work was supported in part by the U.S. Department of Energy (DOE) under Award Number DE-SC0009998. J.H. thanks the Aspen Center for Physics for hospitality and a productive research environment in the early stages of this work.  The Aspen Center is supported by National Science Foundation grant PHY-1607761.  The authors would like to thank Csaba Cs\'aki, Ofri Telem, Simon Catterall, Alex Maloney, Michael Geller, and Sungwoo Hong for helpful discussions as this work was completed.

\bibliographystyle{JHEP}
\bibliography{darkanomalies}

\end{document}